\documentclass[]{emulateapj}
\usepackage{epsfig}
\begin{document}
\topmargin 0.8cm
\newcommand{\etal}{{\it et~al.\/\ }}
\newcommand{\kms}{km~s$^{-1}$}
\newcommand{\Msun}{M$_{\sun}$}

\shorttitle{The Stellar Mass/Halo Mass relation} 
\shortauthors{Munshi et al.}

\title[The Stellar Mass/Halo Mass relation]{Reproducing the Stellar Mass/Halo Mass Relation in Simulated  $\Lambda$CDM Galaxies: Theory vs Observational Estimates   }

\author{Ferah Munshi\altaffilmark{1,2}, F.Governato\altaffilmark{1}, A.M.Brooks\altaffilmark{3}, C.Christensen\altaffilmark{4}, S.Shen\altaffilmark{5}, S.Loebman\altaffilmark{1}, B.Moster\altaffilmark{6}, T.Quinn\altaffilmark{1}, J.Wadsley\altaffilmark{7},
}
\altaffiltext{1}{Astronomy Department, University of Washington, Box 351580, Seattle, WA, 98195-1580}
\altaffiltext{2}{e-mail address: fdm@astro.washington.edu }
\altaffiltext{3}{Astronomy Department, University of Wisconsin, 475 N. Charter St., Madison, WI, 53706}
\altaffiltext{4}{Department of Astronomy, University of Arizona, 933 North Cherry Avenue, Rm. N204, Tucson, AZ 85721-0065}
\altaffiltext{5}{Institute of Particle Physics, University of California, Santa Cruz, CA 95064}
\altaffiltext{6}{Max-Planck-Institute for Astrophysics, Karl-Schwarzschild-Str. 1, 85741 Garching, Germany}
\altaffiltext{7}{Department of Physics and Astronomy, McMaster  University, Hamilton, ON, Canada L8S 4M1}


\begin{abstract} 

  We examine the present--day total stellar-to-halo mass (SHM) ratio
  as a function of halo mass for a new sample of simulated field
  galaxies using fully cosmological, $\Lambda$CDM, high resolution SPH
  + N-Body simulations. These simulations include an explicit
  treatment of metal line cooling, dust and self-shielding, H$_2$ based star
  formation and supernova driven gas outflows. The 18 simulated halos
  have masses ranging from a few times 10$^8$ to nearly 10$^{12}$
  $M_{\odot}$.  At z$=$0 our simulated galaxies have a baryon content
   and morphology typical of field galaxies. Over a stellar
  mass range of 2.2 $\times$ 10$^3$ - 4.5 $\times$ 10$^{10}$
  $M_{\odot}$ we find extremely good agreement between the SHM ratio
  in simulations and the present--day predictions from the statistical
  Abundance Matching Technique presented in \cite{moster12}.  This
  improvement over past simulations is due to a number systematic
  factors, each {\it decreasing} the SHM ratios: 1) gas outflows that
  reduce the overall SF efficiency but allow for the formation of a
  cold gas component 2) estimating the stellar masses of simulated
  galaxies using artificial observations and photometric techniques
  similar to those used in observations and 3) accounting for a
  systematic, up to 30\% overestimate in total halo masses in DM-only
  simulations, due to the neglect of baryon loss over cosmic times.
  Our analysis suggests that stellar
  mass estimates based on photometric magnitudes can underestimate the
  contribution of old stellar populations to the total stellar mass,
  leading to stellar mass errors of up to 50\% for individual
  galaxies.  These results highlight the importance of using proper
  techniques to compare simulations with observations and reduce the
  perceived tension between the star formation efficiency in galaxy
  formation models and in real galaxies.

\end{abstract}




\section{Introduction}
\label{intro}

In the standard $\Lambda$ Cold Dark Matter ($\Lambda$CDM) paradigm
\citep{whiterees78,fall80,blumenthal84, dekelsilk86,whitefrenk91},
many galaxy properties are expected to correlate with the mass of the
galaxy's host halo.  In particular, the stellar-to-halo mass relation
(SHM), defined as the ratio of the stellar mass (M$_{star}$) within a
halo of total mass M$_{halo}$ within a given over-density
($<\rho>/\rho_{crit}$ = 200 in this work) is a robust estimator of
the efficiency of gas cooling and star formation (SF) processes over a
wide range of halo masses \citep{somerville99,bower2010}.  Both
observational \citep{heavens04,zheng07} and theoretical work
\citep{bower11} suggest that the SF efficiency peaks at the scale of
L$^{\star}$ galaxies and declines at smaller and larger masses. On the
low mass end, this decline is likely because SF is suppressed by gas
heating from the UV cosmic field
\citep{quinn96,Gnedin2000,Okamoto2008,nickerson11} and supernova (SN)
heating with gas removal. At larger masses, energy feedback from
super-massive black holes (SMBHs) is thought to be the dominant process responsible for
lowering the SF efficiency
\citep{bower06,croton09,mccarthy11,johansson12}.

Recently, \citet[hereafter M12]{moster12} and other groups
\citep{vale04,conroy06,mandelbaum06,more09,guo10,trujillo11,behroozi12}
used the Abundance Matching Technique (AMT) and its
variations \citep{yang12} to derive a SHM relation of real
galaxies. In its simplest form AMT assumes a monotonic relation
between the stellar mass function of (real) galaxies and the
underlying halo mass function.  This relation is constrained by
matching the observed galaxy stellar mass function to the $\Lambda$CDM
halo mass function from N--body simulations. Similar works
\citep{guo10,leauthaud11,leauthaud12} have included constraints from
lensing and used slightly different underlying cosmologies.  This
approach has also been used to constrain the scatter in the SHM
\citep{reddick12} by comparing the predicted spatial clustering of DM
halos \citep{shethtormen01,reed07, vandaalen12} with the observed
abundances and clustering properties of galaxy populations
\citep{blain04,conroy06,reid10}.  Additionally, \cite{behroozi12} discuss the implications of the upturn in the faint-end slope of the stellar mass function on the SHM relationship.  

Several works have highlighted how uncertainties in the derived SHM
relation depend on a number of factors, some of them poorly known. For
example the stellar masses of real galaxies are inferred from optical
and near--IR photometric measurements and/or resolved spectra
\citep{bell01,kauffmann03b}. This approach carries substantial
uncertainties and possible degeneracies
\citep{bell01,bell03,pforr12,huang12,behroozi10b} as the observed
spectral energy distribution of a galaxy is a function of many
physical processes \citep[e.g., stellar evolution, SFH and
metal-enrichment history, and wavelength-dependent dust attenuation,]
[]{panter04}.  Furthermore, as surveys often measure individual galaxy
magnitudes within an aperture based on a surface brightness cutoff,
the mass of the stellar component could be systematically
underestimated by at least 20\% \citep{graham05,shimasaku01} if part
of it is old (hence faint) and/or low surface brightness. Finally, the
number density of galaxies will be affected by incompleteness at the
faint end of the galaxy luminosity function
\citep{dalcanton98,sawala11,geller12,santini12}.  Separate from worries 
over the stellar mass determinations are worries about the halo mass 
function.   While the halo mass
function obtained in DM-only simulations is robustly constrained
\citep{reed03,millennium05}, it has recently \citep{sawala12} been
shown that halo masses in DM-only simulations exceed those obtained in
simulations including baryon physics by up to 30\%, introducing
another systematic bias in the AMT, as a galaxy of a given stellar
mass is matched with a too massive halo, pushing the stellar-halo mass
ratio down.  Taken together, the above caveats suggest that a better
understanding of the connection between galaxy masses and the
underlying halo masses could in principle be gained by using realistic
simulations of galaxy formation that directly include baryon physics,
SF and SN feedback.

While substantial progress has been made in creating 
galaxies from cosmological initial conditions
\citep{scannapieco10,mccarthy12,sales12,stinson12,johansson12},
several recent studies have pointed out a large discrepancy between
the SHM relation estimated for real galaxies and the one obtained in
several numerical simulations of galaxy formation
\citep{sawala11,guo10}. Simulations have repeatedly shown that a
lack of realistic SN feedback leads to overestimating star formation
as part of the general overcooling problem
\citep{abadi03a,g07,piontek11,keres11}. Most simulations that overproduce
stars form galaxies that have large spheroidal components
\citep{eke01}. Incremental improvements based on more realistic SN
feedback \citep{thacker00,stinson06} led to simulations that formed
galaxies with extended disks \citep{G09,brooks11}, but still
substantially overproduced stars.  Only recently, a new generation of
high resolution simulations demonstrated the impact of feedback at
lowering SF efficiency such that SF occurs only at high gas densities
\citep{ceverino09,G10,eris11,G12,zolotov12,brook12}.  As SF is more
efficient in dense gas clouds, feedback from these high density
regions generate outflows that {\it simultaneously} improve on several
long standing problems namely the substructure overabundance problem
\citep{moore98,klypin99,benson10b}, reducing the B/D ratio in small galaxies by
removal of low angular momentum baryons \citep{binney01,G10,brook11}
and forming DM cores by transferring energy from baryons to the DM
\citep{mashchenko06,Pasetto2010,deSouza2011, Cloet2012, Maccio2012,
  Ogiya2012, pontzen12,teyssier12,G12}.  Forming stars in dense gas
regions is a crucial step, as observations strongly support that the
spatially resolved SF is linked to the local H$_2$ fraction
\citep{bigiel08, krumholz09, genzel12}, which only becomes significant
at the density of star forming regions, $\geq$ 10-100 amu/cm$^3$.

The relationship between gas density and H$_2$ abundance can now be
naturally implemented in simulations, and the creation and destruction
of H$_2$ can be followed consistently \citep{gnedin09, christensen12a}, 
allowing much more realistic simulations to be run, and 
establishes a physically motivated connection between SF and high
density (shielded) gas.  This is indeed one of the major steps
forward in the work presented here.  While feedback from
SNe remains still poorly understood, its effects are being observed
over a large range of redshifts and galaxy masses
\citep{martin99,abel10}.  It is therefore important to evaluate a new set
of high resolution simulations to test if outflows can form galaxies
with realistic observational properties that also reside on the SHM
relation.

Relatively less attention has been given to comparing results
from simulations with observational estimates of the SHM relation in 
a {\it consistent} way. While some recent works reported \citep{sawala11,
avila11,piontek11,Leitner2012} an excess of stars formed in
simulations, they compared the galaxy stellar and total halo
masses directly measured from simulations while those quantities are actually inferred from the light distribution of real galaxies.  
A consistent approach has been already
tried with promising results in \cite{ohsim11}, where two simulated
dwarfs were compared with a set of galaxies from the THINGS survey,
finding that the simulations have the same central baryon and DM
distribution as in the observational sample.  In that work, stellar and
halo masses were obtained using photometric and kinematic data for
both the simulated and the real sample, finding excellent agreement
(see \cite{ohsim11} figure 5).

In this paper, we present a consistent comparison between the SHM
estimated in Moster et al. (2012) and a set of high resolution
simulations spanning 5 orders of magnitude in stellar
masses. The evolution of galaxies is simulated at high resolution using
Smoothed Particle Hydrodynamics (SPH) in a cosmological context. We
focus on how galaxies populate dark matter halos ranging from
$10^8-10^{12} M_{\odot}$ in a field environment and comparing results
with the estimates of the SHM from M12. The four most
massive, MW--like galaxies have similar resolution to the ``Eris''
galaxy \citep{eris11}, while the smaller galaxies have even better force
resolution ( down to 65 pc and star particles as small as 450 M$\odot$). All
simulations include the effects of metal line cooling and H$_2$
dependent star formation.

\begin{figure}[t]
\epsscale{1.2}\plotone{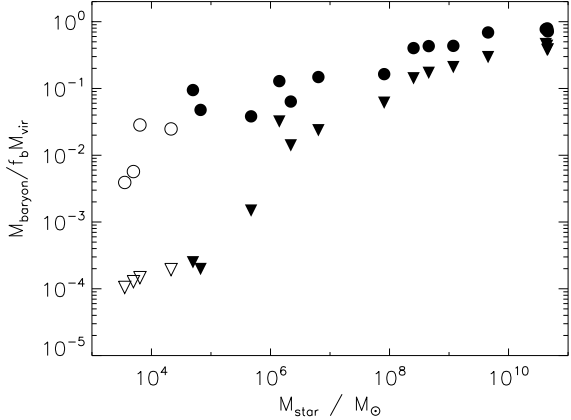}
\caption{ {\it Baryonic fraction with respect to the cosmic ratio, for simulated 
  field galaxies as a function of stellar mass, measured at z=0.}  Circles are the
  ``direct from simulation'' results, including all gas and stars within R$_{200}$.
  Triangles are the ``observable'' baryon fractions, includind all stars and all 
  the `observable' cold gas (defined as 1.4 $\times$ (HI+H2), within R$_{200}$).  
  The empty symbols are galaxies with no observable gas (cold gas 
  mass $<$ 100 M$_{\odot}$).  Galaxies below 10$^{8}$ M$_{\odot}$ lose a 
  significant fraction of baryons due to heating from the cosmic UV background 
  and SN feedback.}
\label{fig1}
\end{figure}

Our simulations include cooling, star formation, a cosmic UV
background and form galaxies with structural properties comparable to
the real ones. The dataset utilized is described in detail in Section 2 and
has been analyzed in other papers showing that the
galaxies follow the Kennicutt-Schmidt law \citep{christensen12a}, have
cored DM profiles similar to the observed ones \citep{ohsim11,G12} and
realistic satellite populations \citep{zolotov12,brooks12}. Without
any further fine tuning, in this work we compare the same simulations
to the SHM relation obtained in M12, using an
analysis technique comparable to the one used in the original paper to
estimate their stellar and halo masses.  We show that simulations form
realistic galactic systems that also match the z$=$0~SHM of real
galaxies over five orders of magnitude in stellar mass.

The paper is organized as follows: in \S2 we describe the details of
our N-body simulations. In \S3 we compare results with the SHM 
predicted in M12. The results are discussed in \S4.

\section{The Simulations}
\label{sims}

The simulations used in this work were run with the N-Body + SPH 
code {\sc GASOLINE}
\citep{wadsley04,stinson06} in a fully cosmological $\Lambda$CDM
context: $\Omega_0=0.26$, $\Lambda$=0.74, $h=0.73$, $\sigma_8$=0.77,
n=0.96. The galaxy sample was selected from two uniform DM-only simulations of
25 and 50 Mpc per side.  From these volumes a few field--like regions
were selected and then resimulated at higher resolution using the
`zoomed-in' volume renormalization technique
\citep{katz93,pontzen08}.  This technique allows for significantly
higher resolution while faithfully capturing the effect of large scale
torques that deliver angular momentum to galaxy halos
\citep{barnes87}. With this approach, the total high resolution sample
contains eighteen field galaxies, each halo resolved by 5$\times$10$^4$ 
to a few 10$^6$ DM particles within R$_{vir}$, defined as the radius at which the average halo density = 200 $\times$
$\rho_{crit}$.

The force spline softening ranges between 64 and 170 pc in the high 
resolution regions of each volume and it is kept fixed in physical 
coordinates at z $<$ 10.  Star particles are formed with a mass of 400-8000
M$_{\odot}$. The halo mass range covered by the simulations spans
nearly four orders of magnitude, from a few times 10$^8$ to 
8 $\times$ 10$^{11}$ M$_{\odot}$ (peak velocities V$_{peak} =$ 10 
to 200 km/sec), and stellar masses M$_{star}$ from 10$^4$ to a 
few 10$^{10}$ M$_{\odot}$. As other
works have highlighted the importance of having a representative sample before
drawing general conclusions \citep{brooks11,sales12,mccarthy12}, the
halos in our sample span a representative range of halo spin values
and accretion histories \citep{geha06}.  
Galaxies and their parent halos were first identified using 
AHF\footnote{{\bf{A}}MIGA's {\bf{H}}alo {\bf{F}}inder, available for 
download at http://popia.ft.uam.es/AHF/Download.html} 
\citep{gill04, knollmann09}.  The total halo
mass (including DM, gas and stars) is defined at a radius R$_{vir}$,
defined as the radius at which the average halo density = 200 $\times$
$\rho_{crit}$, consistent with M12. No sub-halos have been included
in our sample, although the most massive galaxies have a realistic
population of satellites \citep{zolotov12}.  More details on this
dataset and the properties of the satellite population are given in
\citet{G12} and \citet{brooks12}.

\begin{figure}[b]
\epsscale{1.0}\plotone{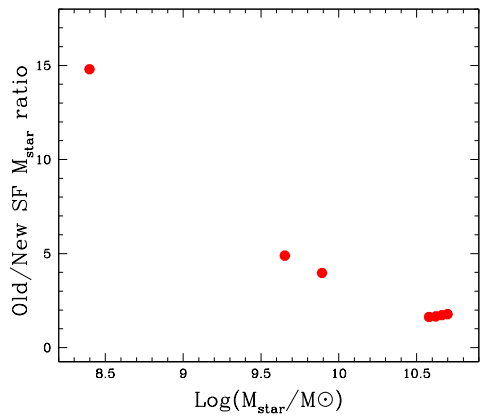}
\caption{{\it The stellar mass ratio between the galaxies
    simulated with the old `low density SF threshold' and the new
    sample}. In the new sample  SF is regulated by the local abundance
    of molecular  hydrogen, resulting in feedback significantly lowering 
    the total SF efficiency.    All quantities as measured directly from 
    the simulations.  }
\label{fig2}
\end{figure}

\subsection{H$_2$ fraction, Star Formation and SN Feedback}

In a significant improvement, this new set of
simulations include metal line cooling \citep{shen10} and a dust
dependent description of H$_2$ creation and destruction by
Lyman-Werner radiation and shielding of HI and H$_2$
\citep[hereafter CH12]{gnedin09,christensen12a}.  As in CH12, the star formation
rate (SFR) in our simulations is set by the local gas density and the
H$_2$ fraction; SF $\propto$ (f$_{H_2}$ $\times \rho_{gas})^{1.5}$.  A
SF efficiency parameter, $c_* =$ 0.1, gives the correct normalization
of the Kennicutt-Schmidt relation (the SF efficiency for each star
forming region is much lower than the implied 10\%, as only a few star
particles are formed before gas is disrupted by SN winds). With the
inclusion of the H$_2$ fraction term \citep[see also][]{kuhlen11}, the
efficiency of SF drops to zero in warm gas with T $>$
3,000 K.  The simulations include a scheme for turbulent mixing that
redistributes heavy elements among gas particles \citep{shen10}.  With this
approach, the {\it local} SF efficiency is linked to the local H$_2$
abundance, as regulated by the gas metallicity and the radiation field
from young stars, without having to resort to simplified approaches
based on a fixed local gas density threshold \citep{G10,kuhlen11}. The
simulations assumed a Kroupa IMF and relative yields, but
observable quantities have been converted to a Chabrier IMF,
for a direct comparison with \cite{moster12}.

\begin{figure}[b]
\epsscale{1.2}\plotone{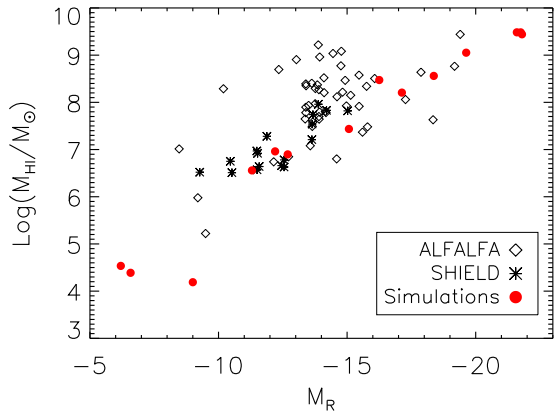}
\caption{{\it The cold gas mass as a function of stellar mass. 
Simulations vs. SHIELD and ALFALFA data.}  
The HI mass of each galaxy in the simulated sample is plotted vs the 
SDSS $r$-band magnitude and compared to two samples from nearby surveys. 
Red solid dots: simulations. Diamonds:
ALFALFA survey. Asterisks: SHIELD survey.  While feedback removes a
large fraction of the primordial baryons, the simulated galaxies have
a high gas/stellar mass ratio, comparable to the observed
samples. Most of the cold gas resides within a few disk scale lengths
from the simulated galaxy centers.}
\label{fig3}
\end{figure}

\begin{figure*}[t]
\epsscale{0.75}\plotone{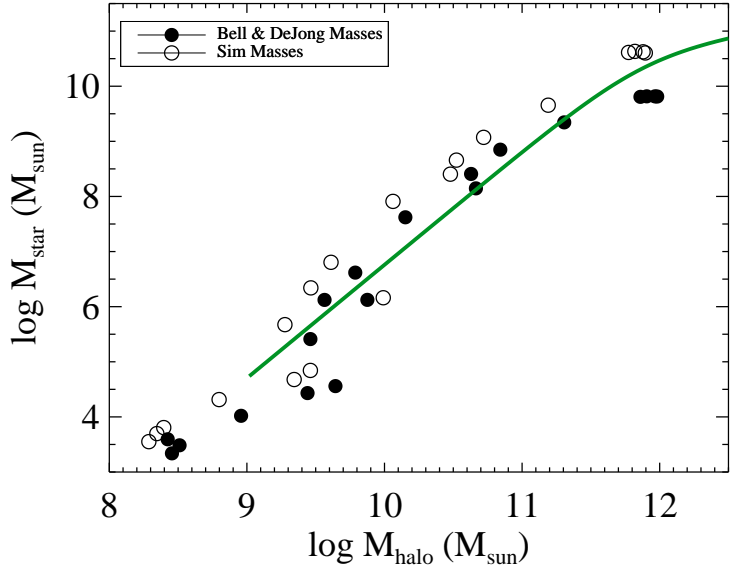}
\caption{{\it The Stellar Mass vs Halo Mass.} 
Black Solid Dots: The SHM relation from our simulations set with 
stellar masses measured using Petrosian 
magnitudes and halo masses from  DM-only runs. This procedure mimics the one followed in 
 M12.  Open Dots: Unbiased stellar masses measured directly from the 
simulations.  Solid Line: Observational results from M12. 
}
\label{fig4}
\end{figure*}

As in previous works using the ``blastwave'' SN feedback approach
\citep{stinson06,G12}, mass, thermal energy, and metals 
are deposited into nearby gas when massive stars evolve into SNe.
The amount of energy deposited amongst those neighbors is 10$^{51}$ 
ergs per SN event.  Gas cooling is then turned off until the end of the 
momentum-conserving phase of the SN blastwave which is set by the local gas density and temperature and by the total amount of energy injected, typically a few
million years. Equilibrium rates are computed from the
photoionization code Cloudy \citep{ferland98}, following \citet{shen10}. 
A spatially uniform, time evolving, cosmic UV background turns on at $z=9$ 
and modifies the ionization and excitation state of the gas, following an 
updated model of \citet{haardtmadau96}. 

This feedback model differs compared to other
``sub-grid'' schemes \citep[e.g.,][]{springel03,scannapieco11} in that
it keeps gas hydrodynamically coupled while in galactic outflows.  The
efficient deposition of SN energy into the ISM, and the modeling of
recurring SN by the Sedov solution, should be interpreted as a scheme
to model the effect of energy deposited in the local
ISM
by {\it all} processes related to young stars, including UV radiation
from massive stars \citep{hopkins11,wise12}. The SFHs of the galaxies
in our simulated sample are bursty, especially those residing in halos
smaller than 10$^{10}$ M$_\odot$.  As discussed in \citet{brook11} and
\citet{pontzen12}, a bursty SFH is necessary to remove low angular
momentum baryons and create the fast outflows able to transfer energy
from baryons to the DM component and create DM cores \citep{G10}.
These processes lead to realistic dwarf galaxies with slowly rising
rotation curves and typical central surface brightnesses 21 $<$
$\mu_{B,o}$ $<$ 23.5 \citep{ohsim11}.  Outflows and the cosmic UV
field progressively suppress star formation in halos with total mass
smaller than a few 10$^{10}$ \Msun. Test runs verify that the
effect of energy feedback on suppressing SF is much larger than that
of having a low  H$_2$ fraction-SF efficiency \citep{christensen12a}.

\subsection{Star formation efficiency as a function of galaxy halo mass}

As a result of SF feedback processes, the SF efficiency is greatly
reduced over the whole mass range of our simulated sample. 
The smallest galaxies in our sample turn only $\sim$0.01\% of their 
primordial baryons into stars. The more massive galaxies in our sample 
turn $\sim$30\% of their primordial baryon content into stars, but 
we demonstrate in the next section that using stellar masses based on 
photometry reduces this efficiency to an apparent 10\%.  Feedback 
expels about 70\% of the gas to outside of R$_{vir}$ in dwarf galaxies 
with v$_c$ $\sim$ 40-50 \kms.  Larger galaxies retain a larger fraction 
of their original baryons, while the smallest galaxies lose an even 
larger fraction of baryons due to gas heating by the cosmic UV 
background, which further reduces their SF (see Figure 1).

To evaluate the effect of the adopted SF model on the resulting SF
efficiency of our galaxy sample, galaxies were re-simulated using the
SF approach used in our older works.  These reference runs adopt a
lower density threshold for SF, 0.1 amu/cm$^3$.  As discussed in
several works \citep{G10,eris11}, a low density threshold makes SF
more diffuse and locally less efficient. 
In this scenario, typical of low resolution simulations where star
forming regions cannot be resolved, the amount of SN energy per unit
mass delivered to gas particles is effectively lowered, making
feedback much less effective at suppressing SF.  While the galaxies in
the low threshold sample have realistic disk sizes \citep{brooks11}
and the total amount of energy released into the gas is actually a few
times larger, they contain many times more stars (see Figure 2) and
overproduce stars compared to the SHM relation. The comparison between
the amount of stars formed in the old and the new runs (Figure 2)
demonstrates that the large decrease in the SF efficiency in the new
simulations (as much as a factor of 15) is due to the improved implementation 
of SF in dense/H$_2$ rich regions \citep[see also][]{G10}. This lower 
SF efficiency goes a long way toward reconciling simulations of galaxy
formation with current estimates of the SHM relation.

\subsection{The Baryon Content of the Simulated Galaxies}

The more massive galaxies in our sample are disk dominated,
transitioning to irregular galaxies below $\sim$10$^{10}$ M$_{\odot}$.
The outflows (and the UV cosmic field in halos below $\sim$10$^9$
M$_{\odot}$)  significantly lower the baryon fraction of the host
halos, with a strong trend of lower baryon fractions at smaller halo
(stellar) mass (see Figure 1).  However, because the fraction of remaining gas turned
into stars is low at all galaxy masses and especially in dwarfs, the
galaxies in our sample have relatively high cold gas to stellar mass
ratios, typical of real galaxies over the tested range. In Figure 3 we
compare the cold gas (HI) content of the simulated sample to the
nearby HI surveys ALFALFA \citep{giovanelli05} and SHIELD
\citep{cannon11}.  The observed dataset includes only galaxies closer
than the Virgo cluster, for a better comparison with our sample of
field galaxies.  With the caveat that selection effects can still play
a role, there is very good agreement between simulations and
observations.
 
The HI masses are directly measured from the simulations, where  HI and H$_2$\footnote{note that H$_2$ masses are small and, while  neglected in Figure 3, contribute little to the overall cold gas mass} abundances are 
calculated on the fly. Magnitudes are measured in the
SDSS $r$-band. 

We verified that the cold gas fraction in the comparison
runs adopting a low density threshold for SF is about a factor of ten
lower at all halo masses.  Lower resolution simulations of small mass
systems have often reported the formation of relatively gas poor
galaxies \citep{g07,colin10,avila11,sawala11}. Low gas content in
simulated low-z galaxies is likely due to a high efficiency of gas to
stars conversion \citep{piontek11} and/or to an excessive loading
factor of the SN winds. 
\section{The Stellar Mass - Halo Mass  relation}

Once individual galaxies in our sample have been identified with AHF,
the Stellar Mass - Halo Mass ratio for each galaxy can then be
obtained. The definition of ``halo mass'' includes all DM and baryons
within an overdensity of 200, but not the mass associated with
individual satellites (a few \% of total at the most). All stars not
in satellites, but within R$_{200}$ are associated with the central
galaxy in the halo. This simple approach is similar to what has been
done in several previous works \citep{sawala11,brook12} and similar to
what has been used in previous comparisons between simulations and the
SHM relation obtained using the AMT \citep{guo10,moster12}.  Our new
sample of simulated field galaxies (open circles in Figure 4)
follows closely the shape and normalization of the present day SHM
relation presented in M12 over the 10$^9$--10$^{12}$ M$\odot$ halo
mass range (solid line), with M$_{star} \propto M_{halo}^2$.  This is
a large improvement over most published works and confirms results on
smaller samples \citep{G10,ohsim11,eris11,brook12} that adopted
similar SF and feedback recipes.

Clearly, an approach where a) SF is limited to dense, H$_2$-rich gas
clouds \citep[a highly correlated situation, see][]{christensen12a}
and b) feedback is hydrodynamically coupled to outflows significantly
reduces the SF efficiency and the present day stellar mass in galaxy
sized halos over a wide mass range.  Our simulations show that both a)
and b) are alone not sufficient, but the combination is sufficient. We have first verified that in
the absence of feedback the SF efficiency remains high even if the
consumption rate of gas is long, as over the course of a Hubble time
most cold gas within the galaxy eventually turns into stars. Moreover,
lack of SN feedback fails to remove the low angular momentum gas,
originating galaxies with an excessive spheroidal component
\citep{G10,brook11}. Similarly, if the identical SN feedback recipe
used in this work is applied to simulations where SF is allowed in
cold, but relatively low density gas (e.g 0.1 amu, as often adopted in
the past), it fails to significantly lower the overall SF efficiency.
In our sample, the mass in stars formed by $z=0$ in galaxies with
H$_2$/high density regulated SF is lower by as much as fifteen
compared to simulations of the same halos adopting a lower density
threshold.  The overproduction of stars in the low threshold runs
occurs {\it even} when  metal lines cooling is neglected.

In summary, while the SF efficiency in the high threshold simulations 
is lower, the cold gas content is similar to that observed in real galaxies 
(see \S 2.3).  Hence, a low SF efficiency was not obtained by simply 
increasing the feedback strength and ``blowing away'' all the baryons. 
Combined, these results show that adopting a more realistic description 
of {\it where} stars form and how feedback regulates SF leads to 
realistic simulations of galaxies.

However, for a meaningful comparison with observations and the SHM 
inferred in M12, it is important to infer {\it both} the stellar and 
halo mass from the simulations using the same techniques as the 
observations.  This additional step is necessary as simulations
directly measure the {\it mass distribution}, while observations infer
the stellar mass from the {\it light distribution}.  Below we show 
that this more accurate approach affects the results substantially.  
We will provide, for the first time, an accurate comparison with the 
present day SHM relation obtained from observational data. We used the
following procedure:

\begin{itemize}

\item{Magnitudes based on the age and metallicity of each star particle 
    were derived using the Starburst99 stellar population synthesis models
    of \citet{Leitherer1999} and \citet{Vazquez2005}, adopting a 
    Kroupa (2001) IMF.  }



\item{For each simulated galaxy, Petrosian aperture magnitudes 
    \citep{blantonetal01,yasuda01} were obtained in the $r$ band.
    This step is necessary as
    observations are limited by the surface brightness of
    the target galaxy dropping below the sky brightness. This
    systematic bias underestimates the amount of light associated with
    individual galaxies, and it is estimated to be of the order of 20\%
    for real galaxies \citep{dalcanton98,blantonetal01}. As our
    galaxies have light profiles that closely mimic those of real
    galaxies \citep{brooks11}, applying this constraint is appropriate. We
    verified that the amount of light lost  is similar to that
    estimated for observational samples. }

\item{The stellar mass of each galaxy was then estimated
    based on its B-V color and V total magnitude, assuming a Salpeter
    IMF to be consistent with adopting the same
    fitting formula as in \citet{bell01}, namely
    $L_V=10^{-(V-4.8)/2.5}$ and then $M_{star}= L_V \times
    10^{-0.734+1.404 \times (B-V)}$.  We then utilize a conversion from Salpeter
    to Chabrier IMF to remain consistent with M12. We find that this procedure systematically {\it
      underestimates} the true stellar masses (by summing all star
    particles within $R_{vir}$ not in satellites) of galaxies by
    20-30\%. This result extends over the whole range of galaxy
    masses.  We find that the specific criteria adopted in
    \cite{bell01} tends to underestimate the contribution from old
    (i.e., high M/L) stellar populations. }

\item{The halo mass for each simulated galaxy was measured re-running
    each simulation as DM--only, matching halos between the two runs
    and counting all mass with R$_{vir}$ (again defined as the radius
    within which the average overdensity is $<\rho> = 200
    \rho_{crit}$). This step is necessary to follow the procedure
    adopted in \cite{moster12}, where halo masses were obtained from a
    large cosmological simulation that did not include gas
    physics. This procedure avoids a subtle, but significant and
    systematic bias between the total halo mass measured in DM-only
    runs vs those of the same halos in simulations that include gas
    physics and feedback. In the latter simulations, feedback can
    remove a significant fraction of the baryons from the halo,
    decreasing the total mass within a fixed physical radius.  The
    virial radius, if defined at a fixed overdensity, then shrinks,
    leading to a smaller total halo mass.  The decrease in M$_{vir}$
    varies with mass, as it depends on the amount of baryons lost, but
    it can be significant, up to 30\%. Since the lowest mass galaxies
    in our sample have lost the most baryons (in winds and UV
    background heating), they can experience a decrease of $\sim$30\%
    in halo mass.  At the high mass end, where galaxies retain most of
    their baryons, the simulated galaxies still see a change of 5-10\%
    in total halo mass compared to the DM-only run (where obviously no
    baryon mass loss is possible). These results are consistent with
    estimates in \citep{sawala12}. \cite{sawala12} also interprets this
    shift as a systematic reduction in the matter infall rate. As even
    small amounts of baryons are removed, the gravitational attraction
    on surrounding material decreases, leading to a decrease in the
    infall rate of both gas and DM and, overtime, to a smaller halo
    mass. Neglecting this effect results in moving the simulations
    datapoints {\it to the left,} away from the SHM relation inferred
    using DM-only runs. This bias is particularly noticeable at small
    galaxy masses, where the SHM relation is steeper.}
\end{itemize}

\begin{figure}[!ht]
\centering
\epsfig{file=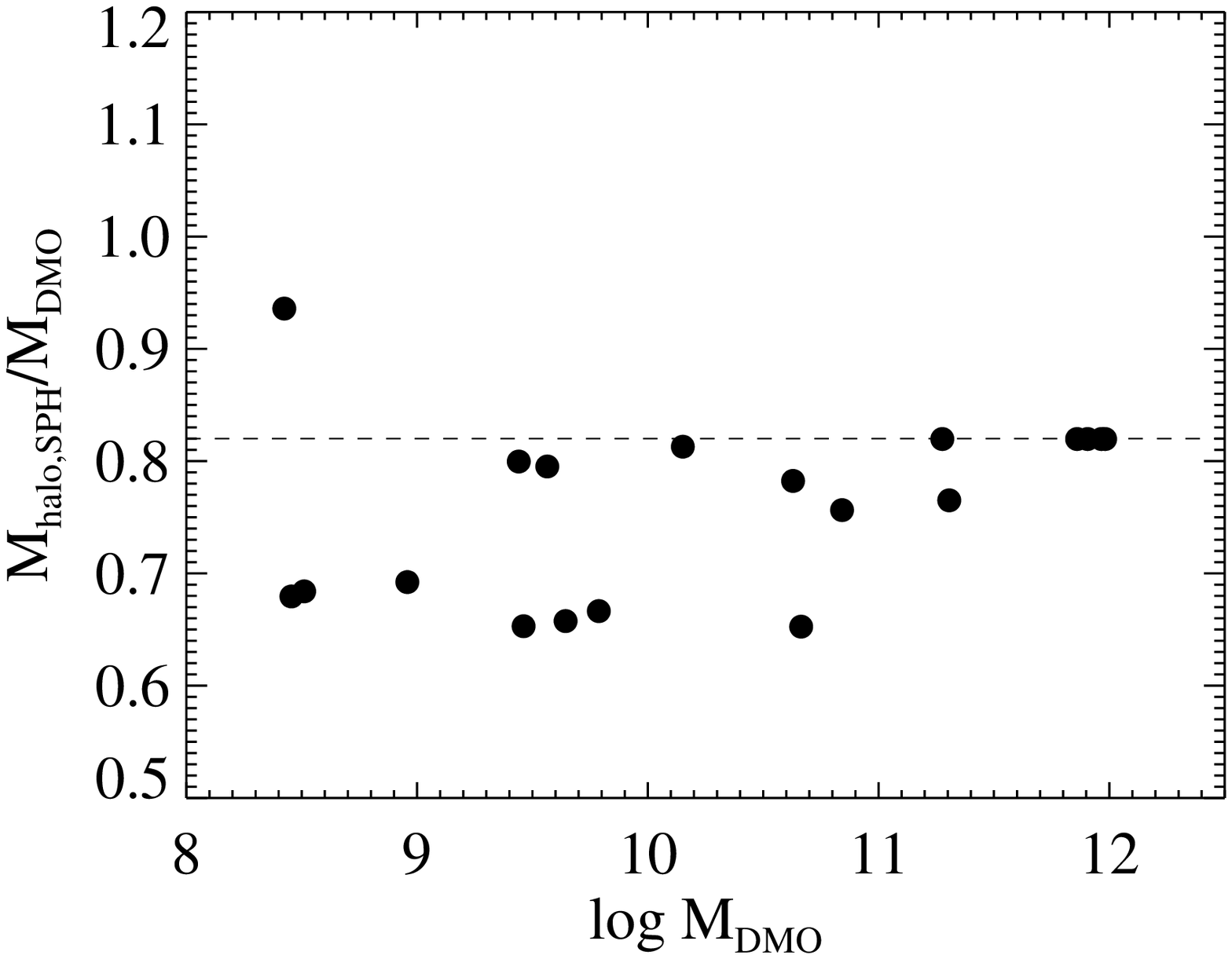, width=0.9\linewidth}
\epsfig{file=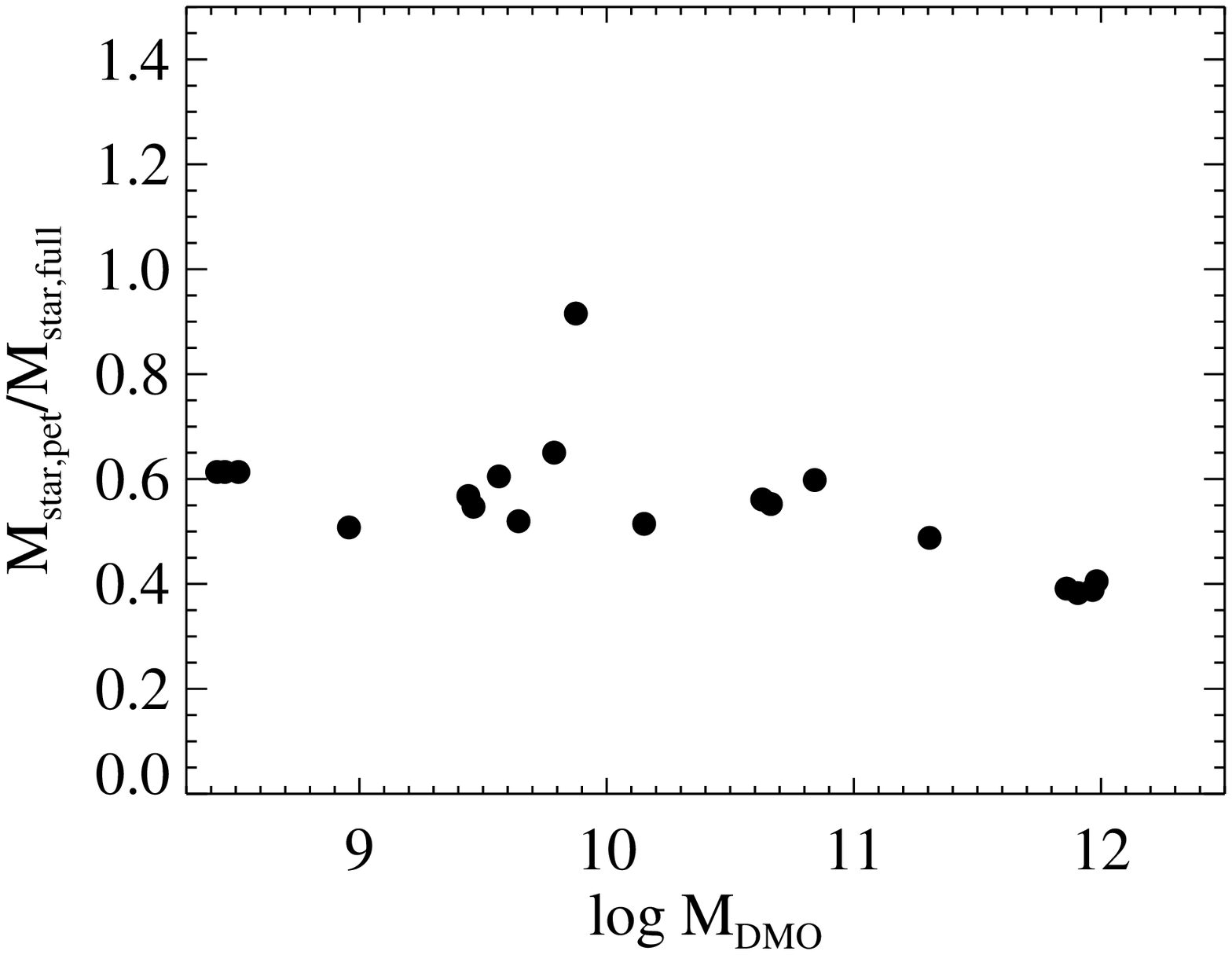, width=0.9\linewidth}
\caption{Top Panel: {\it Halo mass ratio of galaxies in runs with
    baryons and SF vs DM-only runs}.  Individual halos in DM--only
  runs are typically 30\% more massive than their counterparts in
  simulations with gas physics and SF. The effect is smaller in more
  massive halos, where baryon loss due to feedback is less (see also
  \cite{sawala12}). The dashed horizontal line marks the ratio if
  halos had a 100\% baryon loss.  Bottom panel: {\it Estimated vs. True
    Stellar Mass as a function of halo mass.} The stellar mass using
  artificial Petrosian magnitudes and measured using the photometric method in \citep{bell01} vs the ``true'' Stellar mass
  measured directly from the simulations. Stellar masses measured
  using the photometric method in \citep{bell01} in combination with the flux loss from applying the petrosian magnitudes are underestimated by
  about 50\% across the range of galaxy masses in our study.  }
\label{fig5}
\end{figure}

In Figure 4, the black solid dots show results from our simulations
dataset, but using the procedure outlined above, which closely matches
that adopted in M12. The normalization of the SHM is $\sim$
40\% lower than that inferred using the simulations quantities at face
value and closer overall to the SHM relation of M12. There is very
good agreement between the SHM inferred from the `artificial
observations' and the fit in M12 over at least 4 orders of magnitude
in halo mass.  The mass of the brightest galaxies (close in mass
and morphology to Milky Way analogues) goes from being higher to being
slightly {\it lower} than the SHM. This result confirms that the
``blastwave'' feedback implementation is able to reduce the SF
efficiency not only at small halo masses but also in present day halos
around 10$^{12}$ M$\odot$ \citep[see also][]{eris11}.

In Figure 5a we show the total halo mass ratio between the simulations
that include baryons and SF vs the DM-only ones. As discussed above,
halos in DM-only runs are consistently (and significantly, about 30\%)
more massive.  In Figure 5b we show the ratio between the stellar
masses obtained using a combination of Petrosian Magnitudes with \cite{bell01} M/L ratios (closely following M12) and stellar masses derived directly from the simulations.  A systematic
bias of about 50\% is evident across the whole mass range.  The
results from this section highlight the importance of a careful
comparison between simulations and observations. This approach further
reinforces our findings that the modeling of SN feedback greatly reduces 
the tension between the present time SHM relation inferred from the
Abundance Matching Technique for halos with total mass $<$ 10$^{12}$
M$\odot$ and the predictions from hydrodynamical simulations in a
cosmological context. Without fine tuning, this better agreement also
preserves the morphology of the galaxies formed, with disk-dominated
massive galaxies and low mass irregulars that are gas rich.  In future
work, we plan to extend this approach to higher redshifts, using the
appropriate criteria to measure stellar masses in high-z galaxies
\citep[e.g.,][]{pforr12,maraston12}.

\section{Conclusions} 

We have measured the SHM (stellar mass -- halo mass) relation for a
set of field galaxies simulated in a $\Lambda$CDM cosmology and
compared it with the redshift zero predictions based on data from the
SDSS and the Abundance Matching Technique described in M12. The comparison revealed very good agreement in
normalization and shape over five orders of magnitude in stellar mass.
The new simulations include an explicit description of metal lines
cooling and H$_2$ regulated SF, and SN driven outflows.  The
combination of SF driven by the local efficiency of H$_2$ and outflows
reduce the overall SF efficiency over the whole Hubble time by almost
an order of magnitude compared to older simulations, with resulting
M$_{star}$ $\propto$ M$_{halo}^2$. While a large fraction of baryons
is expelled, especially in halos smaller than 10$^{11}$ M$_{\odot}$,
the resulting galaxies have an HI content comparable to those inferred
by local surveys, namely ALFALFA and SHIELD.  The same galaxy set has
a cored central DM density distribution, similar to observations of
real galaxies \citep{G12,brooks12}.

This agreement between simulations and observational data is due to
two systematic factors: 1) An implementation of SF that relates the SF
efficiency to the local H$_2$ abundance in resolved star forming
regions, resulting in localized feedback that significantly lowers the
SF efficiency and 2) ``observing'' the simulations to properly compare
them to observational estimates of the SHM relation.  This approach
involved creating artificial photometric light profiles of the
simulated galaxies and estimating stellar masses based on aperture
magnitudes.  Importantly, it also requires coupling the stellar masses
to halo masses derived from DM-only simulations, rather than the
baryonic simulations.  Our analysis shows that adopting photometric
stellar masses contributes to a 20-30\% {\it systematic} reduction in the
estimated stellar masses.  Stellar mass estimates based on one band
photometric magnitudes are likely to underestimate the contribution of
old stellar populations (reflecting the larger contribution to the
total flux coming from younger stars). This systematic effect is
further exacerbated by the use of aperture based magnitudes, adding
another 20-30\% due to neglecting the contribution of low surface 
brightness populations. Finally, a third systematic effect comes from 
a difference in halo masses in collisionless (DM-only) simulations vs
simulations including baryon physics and outflows. Baryon mass loss
makes halo masses smaller by up to 30\% when calculated at the same
overdensity (200 in our paper and M12). The effect of removing these
biases is to move the simulation points in Figure 4 further lower and
to the right, closer to the SHM.

Notwithstanding the improvements described in this and other recent
works, further adjustment to our numerical schemes to model SF and
feedback processes are most likely  required, as more observational
constraints become available and our understanding of SF improves.  In
future work we plan to extend our analysis of the stellar mass -- halo
mass relation to higher redshifts and larger galaxy masses. The
results presented in M12 point  to a possible discrepancy
between the shape of the star formation history of real galaxies vs
the simulated ones.
 Given the difficulty to obtain robust estimates
from faint and distant galaxy samples, we expect that the approach
outlined in this work, i.e. creating artificial observations to more
directly compare simulations with observations, will play an important
role.

\medskip
\section*{Acknowledgments}

FM, FG and TQ were funded by NSF grant AST-0908499.  FG acknowledges
support from NSF grant AST-0607819 and NASA ATP NNX08AG84G.  AB
acknowledges support from The Grainger Foundation.
Simulations were run at TACC and NAS. We thank Oleg Gnedin, Piero
Madau, Javiera Guedes  for useful discussions.  We thank Jessica 
Rosenberg for ALFALFA data.


\bibliography{bibref}

\bibliographystyle{apj}

\end{document}